\documentclass[runningheads]{llncs}

\usepackage[T1]{fontenc}
\usepackage{graphicx}
\usepackage{booktabs}
\usepackage{xurl}
\usepackage{amsfonts}
\usepackage{dsfont}
\usepackage[table]{xcolor}
\usepackage[hidelinks]{hyperref}

\begin{document}

\title{Improving Cross-Site Whole-Heart Segmentation}
\titlerunning{Appearance Augmentation for CARE-WHS}

\author{
Tanish Mudaliar\inst{1} \and
Justin Li\inst{2} \and
Daniel Lin\inst{3} \and
Julianna Vo\inst{4} \and
Kaitao Liao\inst{5} \and
Xin Wang\inst{6} \and
Shu Hu\inst{1}\thanks{Corresponding author: shuhu@purdue.edu}
}

\authorrunning{Mudaliar et al.}

\institute{
Purdue University, West Lafayette, USA \and
Carmel High School, Carmel, USA \and
Case Western Reserve University, Cleveland, Ohio, USA \and
The Pennsylvania State University, State College, USA \and
Georgia Institute of Technology, Atlanta, USA \and
University at Albany, State University of New York, Albany, USA \\
\email{
\{tmudali,shuhu\}@purdue.edu,
justinyli2018@gmail.com,
dxl941@case.edu,
jhv5097@psu.edu,
kliao47@gatech.edu,
xwang56@albany.edu
}
}

\maketitle  

\begin{abstract}

Whole-heart segmentation from CT and MRI is essential for quantitative cardiac image analysis, but remains challenging under multi-center and multi-modality distribution shift. In the CARE whole-heart segmentation task, models must generalize from limited labeled sites to unseen acquisition distributions, where variation in spacing, intensity, reconstruction texture, and anatomy can degrade out-of-distribution performance. We propose a modality-routed 3D cardiac segmentation pipeline that combines TotalSegmentator-initialized nnU-Netv2 models with site-characterized, label-preserving appearance augmentation. We first characterize the available sites using measurable image properties and use this analysis to motivate candidate data-space generalization routes. The final retained recipe applies Bias Field + Bezier appearance augmentation, combining smooth spatial intensity perturbation with nonlinear intensity remapping, followed by lightweight class-wise largest-connected-component cleanup. On the primary held-out-site validation splits, the final configuration improves CT mean Dice from 0.8350 to 0.9135 and MRI mean Dice from 0.7695 to 0.7830, while also reducing HD95. These results suggest that site-motivated appearance augmentation is a practical strategy for improving cross-site robustness in limited-data whole-heart segmentation. Our code can be found in \href{https://github.com/Purdue-M2/Improving-Cross-Site-Whole-Heart-Segmentation}
{\textbf{\nolinkurl{https://github.com/Purdue-M2/Improving-Cross-Site-Whole-Heart-Segmentation}}}.

\end{abstract}

\section{Introduction}

Three-dimensional cardiac structure segmentation is a core step in quantitative cardiac image analysis. Reliable delineation of chambers, myocardium, and great vessels enables measurements of ventricular volume, myocardial mass and thickness, and vessel morphology, which support assessment of cardiac anatomy and function~\cite{hundley2022scmr}. In clinical and research settings, these quantitative measurements provide objective evidence for disease assessment, longitudinal follow-up, and treatment-response monitoring in cardiovascular conditions such as cardiomyopathy and coronary artery disease~\cite{arbelo2023esc,nieman2024ccta}.

Robust cardiac segmentation remains difficult across heterogeneous data sources. Multi-center data vary with scanner hardware, acquisition protocol, reconstruction, voxel spacing, and patient population, while CT and MRI introduce additional differences in imaging physics and tissue contrast. These factors create distribution shifts that can change both image appearance and apparent anatomy, causing models trained on one site or modality to degrade on unseen acquisition settings~\cite{gao2023bayeseg}.

\begin{figure}[t]
\centering
\includegraphics[width=\textwidth]{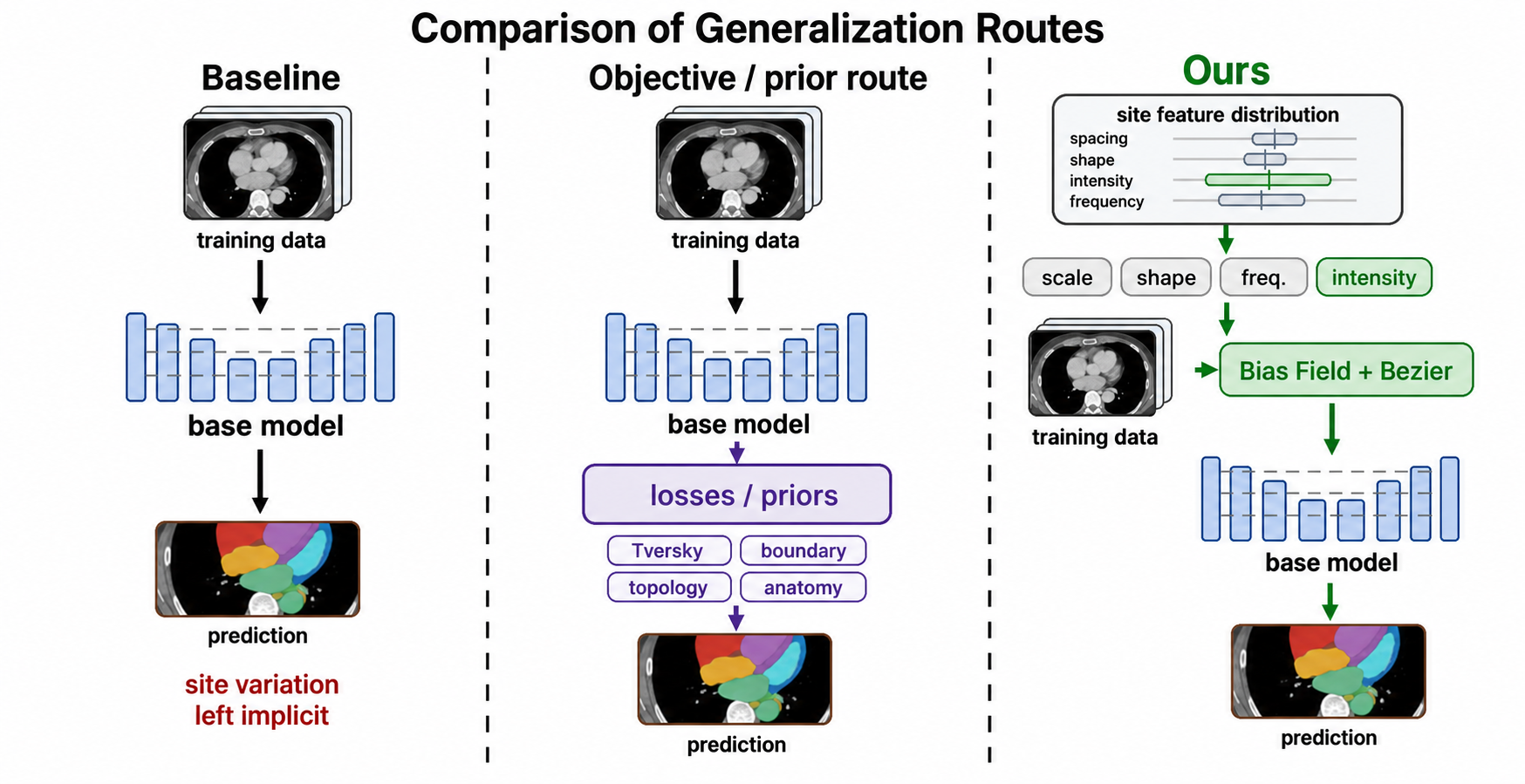}
\caption{Graphical comparison of generalization approaches in model pipeline.}
\label{fig:approach_comparison}
\end{figure}

Existing solutions address generalization at different points in the segmentation pipeline, as summarized in Fig.~\ref{fig:approach_comparison}. Pretrained and generalist models such as TotalSegmentator, MedSAM2, and STU-Net provide strong initialization or baseline model families~\cite{wasserthal2023totalsegmentator,ma2025medsam2,huang2023stunet}. Other methods modify the training signal or impose structural constraints through boundary losses or anatomical priors~\cite{kervadec2021boundary,oktay2018acnn}. Data-based generalization instead acts before the model by expanding the training distribution through label-preserving deformation, contrast, bias-field, or frequency-based transformations~\cite{billot2023synthseg,su2023slaug,ouyang2022gin}.

However, data-space generalization still requires deciding which forms of variation to introduce. Strong initialization may remain sensitive to site-specific appearance or reconstruction cues, while losses and anatomical priors can improve optimization or structural consistency without expanding the image distribution that produced the shift~\cite{kervadec2021boundary,oktay2018acnn}. Generic augmentation can also distort anatomy or remove discriminative image detail when transformations are not well matched to the task~\cite{su2023slaug}. We therefore use site characterization to guide stable data augmentation: measurable site differences motivate candidate augmentation routes, and held-out-site validation determines which components are retained.

In this paper, we apply this strategy to the CARE whole-heart segmentation (CARE-WHS) challenge, where limited CT and MRI training data must generalize to unseen acquisition distributions~\cite{care2026}. We propose a modality-routed 3D cardiac segmentation pipeline that combines TotalSegmentator-initialized nnU-Netv2 models with site-motivated, label-preserving appearance/intensity augmentation. We first characterize the available sites by spacing, physical volume, intensity, and frequency statistics, then use this analysis to motivate candidate augmentation choices. The final retained recipe uses Bias Field + Bezier appearance augmentation and lightweight connected-component cleanup.

\section{Task and Dataset Background}

\subsection{Segmentation Task}
The whole-heart segmentation task evaluates automatic segmentation of cardiac CT and MRI volumes across multiple acquisition sites~\cite{care2026,zhuang2019mmwhs}. Each case is annotated with seven foreground structures: left ventricular myocardium (Myo), left atrium (LA), left ventricular blood pool (LV), right atrium (RA), right ventricular blood pool (RV), aorta (AO), and pulmonary artery (PA). CT and MRI were treated as separate segmentation domains because they differ in acquisition physics, intensity scale, image appearance, and available pretrained initialization~\cite{wang2026unified}.

The available training data contained three CT sites, denoted A, B, and G, and two MRI site groups, denoted C/D and E. The MRI C and D cases were released as a pooled folder without separable site labels, so they were analyzed jointly. To evaluate cross-site generalization, we used held-out-site validation splits throughout development. For CT, the primary split trained on sites A+B and evaluated on held-out site G. For MRI, the primary split trained on C/D and evaluated on held-out site E.

\subsection{Dataset Analysis}
This work uses the CARE-WHS challenge data~\cite{care2026,zhuang2019mmwhs}. Because the challenge evaluates performance on unseen sites, we first characterized the available sites before selecting augmentation and training strategies. Table~\ref{tab:site_characterization} summarizes representative differences in voxel spacing, physical whole-heart volume, median intensity, and low/high-frequency content. These measurements show that site shift is not a single nuisance factor: the CARE-WHS sites differ along scale, appearance/intensity, and reconstruction-texture axes.

\begin{table}
\centering
\fontsize{8}{10}\selectfont
\setlength{\tabcolsep}{3pt}
\caption{Site characterization summary. Spacing is reported as median in-plane spacing \(\times\) through-plane spacing in mm. Whole-heart volume, median intensity, and low/high-frequency ratio are reported as mean \(\pm\) standard deviation.}
\label{tab:site_characterization}
\begin{tabular}{l|l|c|c|c|c|c}
\toprule
Modality & Site & \(n\) & Spacing (mm) & WH Vol. (mL) & Med. intensity & Low/high-freq. ratio \\
\midrule
CT & A & 20 & \(0.43 \times 0.58\) & \(601 \pm 120\) & \(-514 \pm 324\) & \(176.6 \pm 47.5\) \\
CT & B & 20 & \(1.00 \times 1.00\) & \(759 \pm 275\) & \(-106 \pm 161\) & \(19.5 \pm 6.6\) \\
CT & G & 20 & \(0.37 \times 0.50\) & \(533 \pm 118\) & \(-66 \pm 41\) & \(70.6 \pm 11.0\) \\
MRI & C/D & 20 & \(0.94 \times 1.20\) & \(784 \pm 259\) & \(96 \pm 53\) & \(45.8 \pm 14.7\) \\
MRI & E & 26 & \(0.94 \times 1.27\) & \(921 \pm 228\) & \(81 \pm 42\) & \(32.5 \pm 12.7\) \\
\bottomrule
\end{tabular}
\end{table}

This characterization motivated a data-space generalization strategy. Rather than relying only on architecture changes or loss reweighting, we evaluated augmentations that directly expand the training distribution along the measured axes while preserving the underlying anatomical labels~\cite{billot2023synthseg,su2023slaug}. Scale augmentations target spacing and apparent-size variation, deformation-based augmentations target shape variability, and intensity or frequency augmentations target site-dependent appearance and texture~\cite{ouyang2022gin}. The final method was then selected by held-out-site validation rather than by the measured distributions alone.

\section{Methods}

\subsection{Overview}
% \shu{Focus on the framework description for Fig.1}
Given a 3D cardiac CT or MRI volume, our objective is to segment seven whole-heart structures while generalizing across acquisition sites not seen during training. Following the site analysis in Section~2, we treat cross-site variation as a data-distribution problem and use label-preserving appearance augmentation to broaden feature variability without altering anatomical labels.

\begin{figure}[t]
\centering
\includegraphics[width=\textwidth]{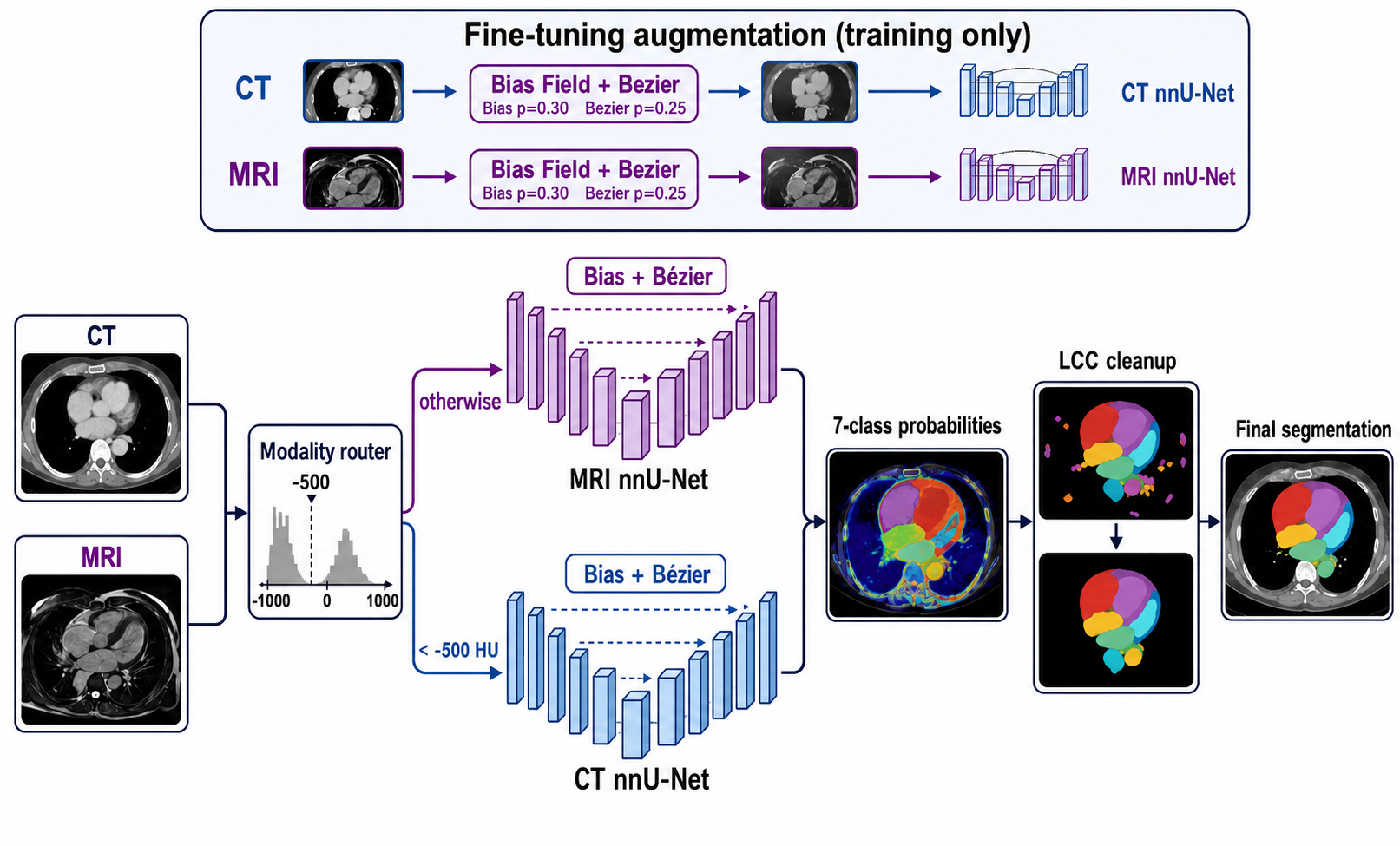}
\caption{Architecture pipeline end-to-end for CT and MRI input streams for training and inference.}
\label{fig:inference_pipeline}
\end{figure}

Figure~\ref{fig:inference_pipeline} shows the proposed framework. The pipeline contains three stages: modality routing, intensity/appearance segmentation, and lightweight post processing. The modality router assigns each input to a CT or MRI branch using a deterministic intensity heuristic. Each branch uses a TotalSegmentator-initialized nnU-Net model trained with the retained Bias Field + Bezier augmentation. The predicted seven-class probability map is then converted to the final segmentation with class-wise largest-connected-component cleanup. 

\subsection{TotalSegmentator nnU-Net Pre-trained Segmentation Models}

Both branches use nnU-Net 3D full-resolution segmentation models~\cite{isensee2021nnunet}. The CT model is initialized from TotalSegmentator-pretrained weights for the heartchambers high-resolution task (Dataset 301)~\cite{wasserthal2023totalsegmentator}, and the MRI model is initialized from TotalSegmentator MRI pretrained weights for Organs Part 1 (Dataset 850)~\cite{dantonoli2025totalsegmri}. Both models are fine-tuned on the corresponding CARE-WHS modality-specific training cases. Default mirroring augmentation is disabled for both CT and MRI training since the label classes are asymmetric.

We use nnU-Net as the backbone because it provides a strong self-configuring baseline for biomedical segmentation while allowing the study to focus on initialization and cross-site generalization strategies rather than new architecture design~\cite{isensee2021nnunet}.

\subsection{Bias-Field and Bezier Intensity Augmentation}

The retained appearance augmentation combines bias-field perturbation and Bezier intensity remapping~\cite{billot2023synthseg,su2023slaug}. Let $I:\Omega\rightarrow \mathbb{R}$ denote a normalized input volume over voxel coordinates $x\in\Omega$. Bias-field augmentation applies a smooth multiplicative field,

\begin{equation}
    I_{\mathrm{bias}}(x)=I(x)\exp(\alpha s(x)),
\end{equation}

where $s(x)$ is a low-frequency random field generated on a coarse grid and smoothly interpolated to image resolution, and $\alpha$ controls the perturbation magnitude. This transformation simulates smooth spatial intensity nonuniformity while preserving the segmentation label map.

Bezier augmentation applies a random nonlinear contrast remapping~\cite{su2023slaug}. Intensities are clipped and rescaled to $[0,1]$, then transformed by a monotone mapping $g:[0,1]\rightarrow[0,1]$ sampled from a cubic Bezier curve. For control points $P_0=(0,0)$, $P_3=(1,1)$, and random internal control points $P_1$ and $P_2$, the curve is

\begin{equation}
    B(t)= (1-t)^3P_0 + 3(1-t)^2tP_1 + 3(1-t)t^2P_2 + t^3P_3,\quad t\in[0,1].
\end{equation}

The horizontal coordinate defines the input intensity and the vertical coordinate defines the transformed intensity. The resulting curve is converted to a monotone lookup table and applied voxel-wise:

\begin{equation}
    I_{\mathrm{bez}}(x)=g(I(x)).
\end{equation}

For the combined transform, the augmented image is

\begin{equation}
    I_{\mathrm{aug}}(x)=g\!\left(I(x)\exp(\alpha s(x))\right),
\end{equation}

with modality-specific clipping and normalization applied according to the preprocessing pipeline. This transform changes image appearance without changing anatomical geometry, making it appropriate for testing whether cross-site performance is limited by contrast and intensity variation.

\subsection{Test-Time Strategies}

We evaluated lightweight test-time strategies after model prediction, including largest-connected-component (LCC) cleanup which focused on the common problem of fragmentation in image segmentation.

For LCC cleanup, each predicted class is processed independently. Given the predicted segmentation $\hat{y}$ and class $c$, we define the binary mask
\begin{equation}
    M_c(x)=\mathds{1}[\hat{y}(x)=c].
\end{equation}
Connected-component labeling with 26-connectivity partitions $M_c$ into disjoint regions $R_1,\ldots,R_n$. We retain the largest component and remove all smaller components:
\begin{equation}
    M_c'(x)=\mathds{1}[x\in R_{i^\ast}], \quad
    i^\ast=\arg\max_i |R_i|.
\end{equation}
This operation uses only graph connectivity and component size; it does not use model probabilities. 

\subsection{Inference Pipeline}
To resolve two separate models, we use a modality router at inference time. The input modality is selected deterministically from the image intensity range. Volumes with minimum intensity below $-500$ are routed to the CT model, while all remaining volumes are routed to the MRI model. This rule separates CT from MRI in the available training data because CT volumes contain air/background values near $-1000$ HU or below, whereas MRI volumes have non-negative background intensities.

The selected nnU-Net branch outputs per-voxel class probabilities for the seven target structures, followed by class-wise LCC cleanup to produce the final segmentation. The full inference pipeline is shown in Fig.~\ref{fig:inference_pipeline}.

% \subsection{Alternative Training and Test-Time Strategies}

% In addition to data-based generalization, we evaluated selected training-based and test-time alternatives to determine whether remaining errors reflected learned class imbalance, poor false-positive/false-negative tradeoffs, or correctable inference artifacts. Training alternatives included fixed Focal-Tversky and adaptive per-class Tversky losses. Test-time alternatives included largest-connected-component cleanup, multi-scale test-time augmentation, registration-based scale correction, and selected class-logit correction experiments.

% These alternatives are reported as ablations rather than as components of the submitted method unless they improve held-out-site validation reliably. This separation keeps the submitted pipeline limited to validated components while still documenting whether training-space or post-hoc strategies explain remaining cross-site failures.

\section{Experiments}

\subsection{Experimental Settings}

\textbf{Datasets}. All experiments use the CARE-WHS data described in Section~2. For development, CT trains on A+B and tests on G and MRI trains on C/D and tests on E. Final submissions are retrained on all available labeled cases for each modality.

\smallskip
\noindent
\textbf{Evaluation Metrics}.
We predominantly employ the Dice similarity coefficient and use 95th-percentile Hausdorff distance (HD95) to measure the performance of our models and baseline methods.

\smallskip
\noindent
\textbf{Baseline Models}.
To contextualize performance under the same held-out-site protocol, we compare our method against SegVol~\cite{du2024segvol}, MedSAM2~\cite{ma2025medsam2}, SAM-Med3D~\cite{wang2025sammed3d}, STU-Net-B~\cite{huang2023stunet}, and nnU-Netv2~\cite{isensee2021nnunet}.

\smallskip
\noindent
\textbf{Baseline Training Details}.
All baselines used the same held-out-site protocol: CT trained on A+B and evaluated on G, while MRI trained on C/D and evaluated on E. SegVol was initialized from \texttt{SegVol\_v1.pth} and fine-tuned for 100 epochs using batch size 1, AdamW, learning rate \(1\times10^{-4}\), weight decay \(1\times10^{-5}\), and its default pseudo-label/semi-supervised setup. MedSAM2 was fine-tuned for 75 epochs with the Hiera-tiny configuration, base learning rate \(5\times10^{-5}\), vision learning rate \(3\times10^{-5}\), cosine decay, gradient clipping, and layer-wise learning-rate decay. SAM-Med3D was evaluated zero-shot from \texttt{sam\_med3d\_turbo.pth} with one click per case. STU-Net-B was initialized from the released \texttt{STU-Net-B.model} checkpoint and fine-tuned with nnU-Netv2 defaults, initial learning rate \(1\times10^{-3}\), batch size 2, and disabled mirroring. The nnU-Netv2 baseline was trained without TotalSegmentator initialization using standard Dice + cross-entropy loss, initial learning rate \(1\times10^{-2}\), polynomial learning-rate decay, and 250/300 epochs for CT/MRI.

\smallskip
\noindent
\textbf{Our Model Implementation Details}. All experiments use TotalSegmentator-pretrained nnU-Netv2 3D full-resolution models. We keep the nnU-Netv2 default training configuration unchanged outside of disabling mirroring. Models are trained with initial learning rate \(1\times10^{-2}\), weight decay \(3\times10^{-5}\), and polynomial learning-rate decay with exponent 0.9. CT models are trained for 250 epochs while MRI models are trained for 300 epochs, each from their respective initialized TotalSegmentator checkpoints. Bias Field augmentation uses magnitude 0.25 with a \(3\times3\times3\) random control-point grid and application probability 0.30. Bezier intensity augmentation uses monotonic ``similar'' mode with lookup-table size 1000 and application probability 0.25.

\subsection{Results}

\textbf{General Performance}.

\begin{table}
\centering
\small
\setlength{\tabcolsep}{2pt}
\caption{Comparison with external baseline models. CT models are trained on sites A+B and evaluated on site G; MRI models are trained on C/D and evaluated on site E. Dice is reported per structure and averaged across foreground classes.}
\label{tab:baseline_comparison}
\begin{tabular}{l|c|c|c|c|c|c|c|c|c}
\toprule
Model & Myo & LA & LV & RA & RV & AO & PA & Dice & HD95 \\
\midrule
\multicolumn{10}{l}{\textit{CT (train: A+B; test: G, \(n=20\))}} \\
\midrule
SegVol      & 0.6234 & 0.9395 & 0.8511 & 0.8276 & 0.7846 & 0.9541 & 0.6403 & 0.8029 & 24.07 \\
MedSAM2     & 0.4180 & 0.9412 & 0.8381 & 0.7252 & 0.5127 & 0.8050 & 0.3330 & 0.6533 & 69.07 \\
SAM-Med3D   & 0.6447 & 0.9103 & 0.6881 & 0.8730 & 0.8712 & 0.8487 & 0.8263 & 0.8089 & 6.21 \\
STU-Net-B   & 0.9141 & 0.8890 & 0.8597 & 0.8858 & 0.4780 & 0.9465 & 0.8670 & 0.8343 & 14.05 \\
nnU-Netv2   & 0.9261 & 0.2540 & 0.8202 & 0.8461 & 0.2027 & 0.6435 & 0.5713 & 0.6091 & 29.13 \\
\rowcolor{gray!20}
Ours        & 0.9477 & 0.9606 & 0.9454 & 0.9115 & 0.7813 & 0.9647 & 0.8834 & \textbf{0.9135} & \textbf{5.22} \\
\midrule
\multicolumn{10}{l}{\textit{MRI (train: C/D; test: E, \(n=26\))}} \\
\midrule
SegVol      & 0.4776 & 0.7299 & 0.7813 & 0.7206 & 0.5148 & 0.5788 & 0.6560 & 0.6370 & 16.49 \\
MedSAM2     & 0.1134 & 0.8087 & 0.8880 & 0.5883 & 0.5623 & 0.5162 & 0.5160 & 0.5704 & 35.86 \\
SAM-Med3D   & 0.5347 & 0.8713 & 0.7669 & 0.8581 & 0.7832 & 0.7238 & 0.7303 & 0.7526 & \textbf{8.41} \\
STU-Net-B   & 0.7494 & 0.8420 & 0.8854 & 0.8172 & 0.7055 & 0.5200 & 0.5002 & 0.7171 & 26.77 \\
nnU-Netv2   & 0.8002 & 0.8879 & 0.9308 & 0.8888 & 0.8439 & 0.5515 & 0.5502 & 0.7790 & 19.26 \\
\rowcolor{gray!20}
Ours        & 0.8082 & 0.8898 & 0.9347 & 0.8974 & 0.8606 & 0.5443 & 0.5460 & \textbf{0.7830} & 19.30 \\
\bottomrule
\end{tabular}
\end{table}

Table~\ref{tab:baseline_comparison} compares the proposed method with external baseline models on the primary held-out-site validation splits. On CT, our method achieves the highest mean Dice, improving over the strongest non-final baseline from 0.8343 to 0.9135. The largest gains are observed for structures that are more sensitive to cross-site appearance and extent variation, including LA, AO, PA, and the overall mean score. Foundation/generalist models such as SegVol and MedSAM2 show competitive performance for selected structures, but their mean performance is lower and less consistent across all seven labels.

On MRI, the margin is smaller but the proposed method again achieves the best mean Dice. The nnU-Netv2 and SAM-Med3D baselines are close in average performance, indicating that MRI site E is less strongly separated by the retained appearance-augmentation route than CT site G. Nevertheless, our final model provides the most consistent overall result, with improvements in Myo, RA, RV, and LV while AO and PA remain challenging across methods.

\smallskip
\noindent
\textbf{Augmented Distribution Analysis}. After selecting Bias+Bezier through held-out-site validation, we measured the same site-characterization statistics under the retained augmentation. This analysis is intended to interpret the selected recipe rather than to serve as a model-selection criterion. Dice and HD95 in Table~\ref{tab:appearance_ablation} determine the retained configuration; Table~\ref{tab:post_augmentation_characterization} shows how Bias+Bezier changes the input distribution seen during training.

\begin{table}
\centering
\small
\setlength{\tabcolsep}{3pt}
\caption{Site characteristics after Bias+Bezier appearance augmentation. Spacing is reported as median in-plane spacing \(\times\) through-plane spacing in mm. Whole-heart volume, median intensity, and low/high-frequency ratio are reported as mean \(\pm\) standard deviation.}
\label{tab:post_augmentation_characterization}
\begin{tabular}{l|l|c|c|c|c}
\toprule
Modality & Site & Spacing (mm) & WH Vol. (mL) & Med. intensity & Low/high-freq. ratio \\
\midrule
CT  & A   & \(0.43 \times 0.58\) & \(594 \pm 142\) & \(-677 \pm 837\) & \(240.8 \pm 161.1\) \\
CT  & B   & \(1.00 \times 1.00\) & \(766 \pm 308\) & \(-354 \pm 381\) & \(97.0 \pm 163.8\) \\
CT  & G   & \(0.37 \times 0.50\) & \(506 \pm 144\) & \(-59 \pm 444\) & \(89.7 \pm 44.6\) \\
MRI & C/D & \(0.94 \times 1.20\) & \(753 \pm 302\) & \(72 \pm 72\) & \(54.0 \pm 30.1\) \\
MRI & E   & \(0.94 \times 1.27\) & \(918 \pm 251\) & \(71 \pm 52\) & \(39.1 \pm 20.9\) \\
\bottomrule
\end{tabular}
\end{table}

Bias+Bezier primarily expands intensity and frequency variability while preserving the geometric spacing of each site. This matches the intended role of the augmentation: broaden appearance variation without changing the anatomical label structure. This improves held-out-site performance for both modalities as indicated from the ablation studies, with the strongest effect observed in CT where the measured appearance shift is larger.

\subsection{Ablation Study}

To validate the role of key components in our framework, we conduct a
series of ablation studies and evaluate on the held-out validation set specified in Datasets in Section 4.1.

\begin{table}
\centering
\fontsize{8}{10}\selectfont
\setlength{\tabcolsep}{2pt}
\caption{\textbf{Ablation Study}. Appearance augmentation and LCC ablation on held-out sites. CT models are trained on sites A+B and evaluated on site G; MRI models are trained on C/D and evaluated on site E. Dice is reported per structure and averaged across foreground classes; HD95 is the mean 95th-percentile Hausdorff distance (mm).}
\label{tab:appearance_ablation}
\begin{tabular}{l|c|c|c|c|c|c|c|c|c}
\toprule
Variant & Myo & LA & LV & RA & RV & AO & PA & Dice & HD95 \\
\midrule
\multicolumn{10}{l}{\textit{CT (train: A+B; test: G, \(n=20\))}} \\
\midrule
Base                   & 0.9183 & 0.9269 & 0.8747 & 0.8708 & 0.5244 & 0.9312 & 0.7991 & 0.8350 & 33.25 \\
Base + LCC             & 0.9229 & 0.9498 & 0.9390 & 0.8868 & 0.5409 & 0.9662 & 0.8044 & 0.8586 & 9.12 \\
Bias                 & 0.9485 & 0.9233 & 0.9404 & 0.9099 & 0.7793 & 0.9270 & 0.8363 & 0.8949 & 13.01 \\
Bezier               & 0.9399 & 0.9103 & 0.9292 & 0.9084 & 0.7460 & 0.9302 & 0.8489 & 0.8876 & 20.15 \\
Bias + Bezier        & 0.9476 & 0.9455 & 0.9419 & 0.9017 & 0.7727 & 0.9499 & 0.8656 & 0.9036 & 13.66 \\
\rowcolor{gray!20}
Bias + Bezier + LCC  & 0.9477 & 0.9606 & 0.9454 & 0.9115 & 0.7813 & 0.9647 & 0.8834 & \textbf{0.9135} & \textbf{5.22} \\
\midrule
\multicolumn{10}{l}{\textit{MRI (train: C/D; test: E, \(n=26\))}} \\
\midrule
Base                   & 0.7830 & 0.8881 & 0.9302 & 0.8909 & 0.8207 & 0.5252 & 0.5481 & 0.7695 & 24.50 \\
Base + LCC             & 0.7910 & 0.8885 & 0.9314 & 0.8933 & 0.8366 & 0.5106 & 0.5492 & 0.7715 & 20.13 \\
Bezier               & 0.8025 & 0.8919 & 0.9332 & 0.8936 & 0.8510 & 0.5297 & 0.5455 & 0.7782 & 23.23 \\
Bias                 & 0.7990 & 0.8887 & 0.9328 & 0.8828 & 0.8439 & 0.5294 & 0.5419 & 0.7740 & 21.83 \\
Bias + Bezier        & 0.8081 & 0.8898 & 0.9347 & 0.8929 & 0.8582 & 0.5356 & 0.5457 & 0.7807 & 20.87 \\
\rowcolor{gray!20}
Bias + Bezier + LCC  & 0.8082 & 0.8898 & 0.9347 & 0.8974 & 0.8606 & 0.5443 & 0.5460 & \textbf{0.7830} & \textbf{19.30} \\
\bottomrule
\end{tabular}
\end{table}

\textbf{Effect of Appearance Augmentation.}
Table~\ref{tab:appearance_ablation} evaluates the contribution of the retained appearance components under held-out-site validation. Bias Field and Bezier augmentation are tested independently and jointly before adding LCC cleanup. On CT, both appearance transforms improve over the base model, with Bias Field producing the stronger single-transform gain and Bias Field+Bezier giving the best pre-post-processing result. On MRI, the same ordering is less pronounced but the combined recipe remains the strongest augmentation-only setting. These results support the use of paired smooth spatial intensity perturbation and nonlinear intensity remapping rather than either transform alone.

\textbf{Effect of LCC Cleanup.} LCC cleanup is evaluated both on the base predictions and on the final augmented model. Its main effect is reflected in HD95, where removing small disconnected components substantially reduces boundary outliers, with an improvement especially clear for CT. Because LCC improves the final mean Dice in both modalities, it is retained as a lightweight inference-time cleanup step.

\section{Conclusion}

We presented a modality-routed 3D whole-heart segmentation pipeline for cross-site CT and MRI segmentation in CARE-WHS. The method uses TotalSegmentator initialized nnU-Netv2 models with Bias Field + Bezier appearance augmentation chosen from characterizing measurable site variation and appended with a class-wise LCC cleanup.

Across the primary held-out-site splits, the retained recipe improved over the base models for both modalities, with the largest gain observed for CT. The results suggest that label-preserving appearance augmentation (derived from the feature distribution) can be an effective way to improve robustness when limited cardiac training data contain site-specific acquisition differences. LCC cleanup further reduced boundary outliers, particularly in CT, while MRI improvements were more modest.

\textbf{Limitations}.
The site-characterization analysis is specific to the available CARE-WHS training data.
% , and we do not claim that the observed CT or MRI feature distributions generalize unchanged to all cardiac datasets. 
The broader contribution is a development workflow that quantifies site variation and selects candidate augmentation axes that increase feature variance while preserving anatomical labels, retaining components only when they improve held-out-site validation.

\textbf{Future Work} should extend the site-characterization strategy beyond input-space augmentation. Measured site differences could be used to guide feature-space regularization, encouraging the model to learn representations that remain predictive of cardiac anatomy while being less predictive of acquisition site. This may improve generalization to unseen scanners and protocols beyond what label-preserving augmentation alone can provide.

\textbf{Acknowledgments}. This work is supported by the U.S. National Science Foundation (NSF) under grants IIS-2434967 and CNS-2431725, the National Artificial Intelligence Research Resource (NAIRR) Pilot and TACC, Computing Research Association (CRA), Purdue ACC Curricular Innovation Seed Grant, and Purdue Applied AI Research Center. The views, opinions, and/or findings expressed are those of the author and should not be interpreted as representing the official views of NSF, NAIRR Pilot, CRA, and Purdue University or any of the other institutions listed.

\textbf{Disclosure of Interest}. The authors have no competing interests to declare that are relevant to the content of this article.

\end{document}